IEEE*Access*

## RESEARCH ARTICLE

# Robust Pareto Transistor Sizing of GaN HEMTs for Millimeter-Wave Applications

**RAFAEL PEREZ MARTINEZ**, (Graduate Student Member, IEEE),
**STEPHEN BOYD**, (Life Fellow, IEEE), AND SRABANTI CHOWDHURY, (Fellow, IEEE)
Department of Electrical Engineering, Stanford University, Stanford, CA 94305, USA
Corresponding author: Rafael Perez Martinez (rafapm@alumni.stanford.edu)

This work was supported in part by Stanford Graduate Fellowship (SGF).

**ABSTRACT** This paper introduces a robust Pareto design approach for transistor sizing of Gallium Nitride (GaN) High Electron Mobility Transistors (HEMTs), particularly for power amplifier (PA) and low-noise amplifier (LNA) designs in 5G applications. We consider five key design variables and two settings (PAs and LNAs) where we have multiple objectives. We assess designs based on three critical objectives, evaluating each by its worst-case performance across a range of Gate-Source Voltages ($V_{GS}$). We conduct simulations across a range of $V_{GS}$ values to ensure a thorough and robust analysis. For PAs, the optimization goals are to maximize the worst-case modulated average output power ($P_{out,avg}$) and power-added efficiency (PAE$_{avg}$) while minimizing the worst-case average junction temperature ($T_{j,avg}$) under a modulated 64-QAM signal stimulus. In contrast, for LNAs, the focus is on maximizing the worst-case maximum oscillation frequency ($f_{max}$) and Gain, and minimizing the worst-case minimum noise figure (NF$_{min}$). We utilize a derivative-free optimization method to effectively identify robust Pareto optimal device designs. This approach enhances our comprehension of the trade-off space, facilitating more informed decision-making. Furthermore, this method is general across different applications. Although it does not guarantee a globally optimal design, we demonstrate its effectiveness in GaN transistor sizing. The primary advantage of this method is that it enables the attainment of near-optimal or even optimal designs with just a fraction of the simulations required for an exhaustive full-grid search.

## I. INTRODUCTION

Gallium Nitride (GaN) high-electron-mobility transistors (HEMTs) have gained significant attention at millimeter-wave (mm-wave) frequencies thanks to their superior high-power and low-noise performance compared to other III-V and silicon-based semiconductor technologies. These devices are primarily utilized in power amplifiers (PAs), where GaN's intrinsic material properties enable high output power ($P_{out}$) and power-added-efficiency (PAE) across the mm-wave spectrum, which is vital for RF systems in current and emerging wireless links such as 5G and beyond-5G [1], [2], [3], [4]. While their

The associate editor coordinating the review of this manuscript and approving it for publication was Pedro Miguel Cabral.

predominant use is in PAs, GaN HEMTs also demonstrate competitive low-noise performance along with superior breakdown voltage and ruggedness when compared to other III-V high-performance semiconductor technologies (e.g., indium phosphide and gallium arsenide). These attributes enable lower noise amplification and the ability to handle higher input power levels, making GaN HEMTs particularly advantageous for 5G-and-beyond and defense applications. Typically, LNAs based on semiconductor technologies with lower breakdown voltages require protection circuits, such as limiters, to safeguard the amplifier, which increases the system's overall noise figure (NF). GaN-based LNAs streamline system-level designs by reducing complexity and size while maintaining acceptable NF values [5], [6].

Even with the adoption of a high-performance semiconductor technology like GaN, the development of PAs and LNAs remains challenging due to the multifaceted design space, which involves numerous trade-offs. As an example, in the context of 5G technology and PA design, spectrally efficient modulation schemes such as orthogonal frequency-division multiplexing (OFDM), higher-order quadrature amplitude modulations (QAMs), and carrier aggregations are employed to maximize the throughput of the communication channels. In order to maintain a high signal fidelity, stringent linearity standards are imposed in the form of adjacent channel power ratio (ACPR) and error vector magnitude (EVM). As such, PAs are forced to operate below their peak output power ($P_{out}$) level to comply with these requirements, which significantly reduces the modulated average output power ($P_{out,avg}$) and average amplifier efficiency (PAE$_{avg}$) compared to their peak continuous wave (CW) performance. This problem is further exacerbated by device-level issues such as soft compression, dc-RF dispersion, and self-heating, which reduce the device's output power, PAE, and linearity [3], [7]. In particular, as self-heating effects intensify, the device's junction temperature ($T_j$) rises, leading to reduced channel mobility and, consequently, a lower drain current ($I_D$). This sets a boundary on the maximum achievable $P_{out}$, PAE, and the overall operational lifetime of the device. From a reliability standpoint, prior research using the same GaN process as in this study has demonstrated that reducing $T_j$ by 40 °C significantly enhanced the mean time to failure (MTTF) in RF accelerated life tests by an order of magnitude [8].

In LNA design, a direct trade-off exists between the minimum achievable NF and maximum Gain [9]. As one of the initial stages in the receiver, the NF of the LNA must be minimized since the noise contribution of each subsequent stage is less significant as the Gain of the first stage minimizes the noise contribution of the proceeding stages, as suggested by Friis' equation [10]. On top of that, there is also a trade-off between power consumption and LNA performance (e.g., Gain, NF, and linearity), as reducing power consumption generally comes at the expense of lower performance. However, linearity is not typically the main concern in LNAs since the linearity limitations for receivers are more prominent at subsequent stages of the receiver. It is these later stages that primarily determine the receiver's overall third-order-intercept point (IP$_3$) or 1-dB compression point ($P_{1dB}$). An exception to this is in wideband receivers, which might encounter numerous strong interfering signals [11].

Given the previously described challenges, a major concern for circuit designers is choosing the optimal device size for their intended application. Nevertheless, transistor sizing is a complex multi-objective optimization problem in mm-wave circuit design, typically tackled through iterative trial-and-error methods, rarely leading to a definitive solution. In many instances, the device is optimized without considering the whole design space. Some examples include only maximizing the device for the highest peak unity power Gain frequency ($f_{max}$), $P_{out}$, or PAE performance [12], [13]. To find the optimal device size, it is important to evaluate the whole design space by taking into consideration the proper metrics for various trade-offs in a PA or LNA. In the present work, we propose a robust Pareto design approach to size GaN HEMTs by evaluating the ''worst-case'' performance obtained from both LNA and PA metrics and then finding the Pareto optimal designs (i.e., a design that is equally good or better across all objectives and superior in at least one). We utilize a lightweight hyperparameter optimization (HPO) software named Hyperparameter Optimizer: Light-weight and Asynchronous (HOLA) [14] to aid this approach, which searches and retrieves Pareto optimal points with respect to multiple objectives using derivative-free optimization (DFO). Our DFO method achieves near-optimal or even optimal designs with significantly fewer simulations than a full-grid search.

Compared to previous multi-objective optimization studies done in various semiconductor technologies, such as in 2D-material-based FETs [15], stacked nanosheet transistors [16], and thermal/mechanical behavior of GaN HEMTs [17], the goal of our approach is to create device designs that perform reliably in a carefully controlled environment and across a wide range of conditions in mass production. This approach also differs from other Pareto optimization studies, such as the one in [18], which has been conducted for a DC generation system using GaN power devices. We primarily focus on robust Pareto optimization to achieve our goal. This systematic design methodology aims to identify system/process settings that remain robust against variability and uncertainty, optimizing for worst-case scenarios across multiple objectives.

This article is organized as follows. Section II describes and formulates the robust optimization approach to sizing GaN HEMTs. In Section III, we use the described Pareto optimization approach by first evaluating modulated large-signal metrics based on a 64-QAM modulation scheme in the context of PA design, followed by the evaluation of small-signal metrics in GaN HEMTs for LNA design. Section IV describes our derivative-free approach to retrieving Pareto optimal points using HOLA. Lastly, Section V concludes this article.

## II. ROBUST OPTIMIZATION
Robust optimization is a specialized area within the broader field of optimization. It primarily focuses on developing methods and approaches to manage (or reduce) the adverse implications of parameter uncertainty [19], [20], [21], [22]. Within robust optimization, two approaches are prevalent: statistical and worst-case deterministic. The statistical approach models parameter uncertainty as random variables, aiming to optimize the expected value of the objective under the distribution of these variables. Conversely, the worst-case deterministic approach, which is the core of the present work, assumes a range of potential parameter values and focuses on optimizing the desired outcome under the most unfavorable conditions [23]. Analogous to this, in machine learning (ML), this concept is referred to as *adversarial*, which pertains to a method of evaluation where an "adversary" specifically selects the most challenging or worst-case parameter for operational use and then critically assesses the system by examining its performance. This approach is critical to ensuring the resilience and reliability of ML models, particularly against inputs specifically crafted to test their limits, i.e., *adversarial examples* [24]. Consequently, the term *adversarial optimization* is often used interchangeably with worst-case robust optimization.

In engineering design, the principle of worst-case robust optimization transcends beyond optimizing for performance metrics under the most unfavorable conditions. It addresses common oversights in device designs, such as failing to account for known and unknown factors during device usage. For example, some elements of a device's functionality might be predictable, while others remain uncertain. This uncertainty is particularly relevant in manufacturing, where parameters might vary. Therefore, it is crucial that the design works well, not just for a hand-crafted device but also for devices produced on a large scale. Another oversight is assuming complete knowledge about how the device will be used. During the design phase, it is common to calibrate the device based on known factors, but this can introduce biases. Nonetheless, device parameters can be inherently unpredictable, and this variability worsens during the manufacturing phase.

In the context of transistor sizing, there are numerous benefits to employing a robust Pareto design methodology. A robust design results in reduced sensitivity, i.e., the designs are less sensitive to manufacturing and operational variations, such as threshold voltage shifts or quiescent point changes during large-signal operation. In addition to these benefits, identifying Pareto designs offers two practical applications that enhance the decision-making process. The first application allows finding a dense set that characterizes the actual trade-off surface, facilitating intelligent discussions about what is achievable. The second application involves identifying a small representative set of designs, enabling designers to choose from various options with a specific metric in mind. This could be a device with high $P_{out}$, PAE, linearity, or a combination of the three, thereby providing an informed approach to selecting the most suitable design based on the application requirements.

It is essential to clarify that the concepts of Pareto optimality (multi-objective optimization) and robustness are distinct and address different aspects of device design. Pareto optimality ensures we can trade-off between multiple objectives rather than solely optimizing for one. In contrast, robust optimization ensures that the performance of designs remains satisfactory under variations in manufacturing or operational conditions. Users can opt for either approach, e.g., they might choose to implement only Pareto optimality without incorporating robustness, or vice versa. With this understanding of Pareto optimality and robustness, we proceed to our main objective, which integrates both elements. In the present work, our goal is to identify robust Pareto optimal designs that comprehensively explore the trade-off surface. This approach merges the two concepts discussed above: achieving robustness in design to ensure consistent performance under varying conditions and identifying Pareto optimal points to effectively navigate the trade-offs among multiple objectives.

## A. ROBUST OPTIMIZATION PROBLEM FORMULATION
In the present work, we encounter an optimization problem with $k$ objectives $f_1(x, p), \ldots, f_k(x, p)$, where $x$ corresponds to a vector of design variables with $a$ variables, and $p$ is the operating parameter with dimension $b$. Each objective has a "direction," i.e., we want to minimize or maximize it. We also adjust the objectives we wish to maximize by multiplying them by $-1$. Therefore, when considering a multi-objective optimization problem, our goal is to minimize all $f_i(x, p)$.

Given that objectives depend on operating parameters and other parameters not under our control, the robustness of the design is achieved by considering the worst-case scenarios across a plausible range of values for $p$, defined within the set $\mathcal{P}$. Since we are minimizing the objectives, robustness is ensured by taking the maximum value of $f_i(x, p)$ over the set $\mathcal{P}$. We express this mathematically as

$$F_i(x) = \max_{p \in \mathcal{P}} \quad f_i(x, p). \tag{1}$$

In this framework, one design is said to dominate another if it is equally good or better across all objectives and superior in at least one. A design that is not dominated is called *Pareto optimal*. The set of Pareto optimal designs is called the optimal trade-off curve (with two objectives) or surface (with more than two). When considering a problem with $k$ objectives, we restrict our considerations to Pareto optimal points, as accepting any solution that is not Pareto optimal is inherently suboptimal [23].

We consider two settings: PA and LNA transistor sizing. In both settings, we address five design variables (i.e., $a = 5$), which correspond to factors we have control over, particularly the drain-source voltage ($V_{DS}$), finger width ($W_f$), number of fingers ($N_f$), gate-drain–gate spacing (GDG), and gate-source–gate spacing (GSG). Additionally, we have one operating parameter (i.e., $b = 1$), the gate-source voltage ($V_{GS}$). In this context, $V_{GS}$ was chosen as an operating parameter (to not be confused with the transistor's operating condition) as any small changes in $V_{GS}$ can have a significant impact on the device's performance. On the other hand, $V_{DS}$ was selected as a design variable, as it is often a specified requirement provided to a circuit designer.

For PAs, we aim to optimize three objectives (i.e., $k = 3$), which include maximizing the worst-case $P_{out,avg}$ ($P_{out,avg,wc}$) and worst-case $PAE_{avg}$ ($PAE_{avg,wc}$), while minimizing the worst-case average junction temperature ($T_{j,avg,wc}$) under a modulated signal stimulus. Similarly, for LNAs, we also focus on three objectives (i.e., $k = 3$), namely maximizing the worst-case maximum oscillation frequency ($f_{max,wc}$) and worst-case Gain (Gain$_{wc}$), and minimizing the worst-case minimum noise figure (NF$_{min,wc}$). For both settings (e.g., PA and LNA transistor sizing), the worst-case objectives are determined by evaluating their performance across a range of $V_{GS}$ and selecting the values that result in the worst-case performance for each objective.

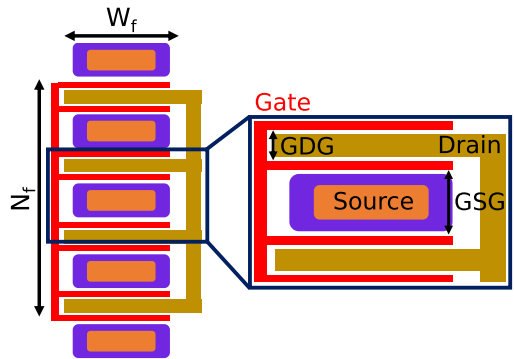

**FIGURE 1.** Qualitative layout of a HEMT device with its corresponding design variables.

**TABLE 1.** Operating parameter range for PA optimization.

| Parameter | Range | Step | # Points |
|-----------|-------|------|----------|
| $V_{GS}$ (V) | $[-1.8, -1.2]$ | 0.1 | 7 |

**TABLE 2.** Design variables for PA optimization.

| Variable | Type | Range | Step | # Points |
|----------|------|-------|------|----------|
| $V_{DS}$ (V) | Float | $[16, 28]$ | 4 | 4 |
| $N_f$ | Even Integer | $[2, 8]$ | 2 | 4 |
| $W_f$ ($\mu m$) | Float | $[25, 100]$ | 12.5 | 7 |
| GDG ($\mu m$) | Float | $[22, 52]$ | 30 | 2 |
| GSG ($\mu m$) | Float | $[33, 78]$ | 30 | 3 |

## III. PA AND LNA OPTIMIZATION

The robust Pareto design methodology is studied using a 150-nm gate length ($L_G = 150$ nm) GaN-on-SiC HEMT from MACOM (previously Wolfspeed) [25] by carrying out thousands of SPICE simulations using the Cree/Wolfspeed/MACOM HEMT model within Keysight Advanced Design System (ADS) [26], [27], [28]. This HEMT model has undergone thorough validation by the foundry across various frequencies, bias conditions, and geometries to ensure good simulation accuracy for the devices utilized in this study. Furthermore, the HEMT model has been validated against measured S-parameter, load-pull, and Gain compression data to verify the model's capability for accurate small- and large-signal simulations. Although the models have been validated through single-tone frequency large-signal measurements, they offer a reliable first-order representation of performance under a large-signal stimulus. Therefore, this work primarily focuses on establishing a workflow for transistor sizing. If the model is improved, simulations can be repeated following the proposed approach.

### A. PA TRANSISTOR SIZING DESIGN VARIABLES AND OBJECTIVES

In the context of PA transistor sizing, we have one operating parameter, $V_{GS}$, and five design variables (i.e., factors we have control over), specifically $V_{DS}$, $W_f$, $N_f$, GDG, and GSG. Table 1 outlines the operating parameter range (with linear spacing) corresponding to $V_{GS}$ utilized in the context of PA transistor sizing, representing the factor chosen by the adversary to assess the worst-case scenario for each objective.

Here, the range was chosen to be across a reasonable bias range where the devices might be typically operated in [8], [29], [30]. It translates to an $I_D$ range of $30 - 275$ mA/mm (for reference, the threshold voltage of the devices is approximately $-2$ V, and this range corresponds to Class AB to Class A operation). We also did not consider bias conditions below 10 mA/mm as current densities below this threshold did not adhere to the foundry model guidelines (i.e., Class B operation was not considered). In practice, the range for robust optimization is primarily driven by the application, and it is up to the user to carefully choose a reasonable range to ensure no additional performance is unnecessarily "thrown away." Furthermore, Table 2 summarizes the type, range, step, and number of points (with linear spacing) for these five design variables, representing a total of 672 designs. In addition, Fig. 1 showcases a qualitative layout of a HEMT, visually showing the physical representation of the design variables used.

With regard to the objectives used for PA transistor sizing, we primarily considered modulation performance metrics as they are a better representation of system-level performance [7]. Interested readers may refer to [7] for a more detailed comparison of various metrics used to quantify GaN HEMT devices in the context of PA applications.

The modulated performance was quantified by a 5G New Radio (NR) uplink Cyclic Prefix-Orthogonal Frequency Division Multiplexing (CP-OFDM) 64-QAM modulation scheme, which was generated using the virtual test bench (VTB) in Keysight ADS. In this modulation scheme, the maximum allowable EVM is 8%, and the maximum allowable ACPR is $-30$ dBc, in accordance with the standards for NR devices [31]. The modulated signal was then utilized to characterize signal distortion, employing methods such as the compact test signal and distortion EVM [32]. This approach was designed to expedite simulation time while maintaining high accuracy, which is necessary to simulate the 672 designs.

In our approach, we have implemented self-heating functionality to capture electro-thermal effects accurately, which are inherently layout-dependent. This was done using Keysight's Electro-Thermal simulator to calibrate self-heating within the MACOM HEMT model for the various geometries. Similarly, each of the device's optimal load-pull impedances that minimized Gain compression while providing high $P_{out}$ and PAE was extracted for the 672 designs (i.e., a Pareto optimal load that maximized $P_{out}$ and PAE while minimizing Gain compression was chosen accordingly). To focus solely on the contribution of the fundamental harmonic, the terminations for all other harmonics (i.e., second, third, fourth, and fifth harmonics) were "opened" (i.e., set to a high value) accordingly.

A frequency of 30 GHz was selected, primarily aiming to align with the 5G frequency bands. The employed modulation scheme is a CP-OFDM 64-QAM signal, with a center frequency ($f_c$) of 30 GHz and a bandwidth (BW) of 10 MHz. The finger width in the designs that were taken into consideration was optimized up from 25 $\mu$m (the minimum value feasible in the model, as imposed by the foundry's restrictions or Design Rule Checking (DRC) flags) to 100 $\mu$m since, beyond this value, there is a noticeable decline in the device's Gain. In a similar manner, the range of the other design variables was chosen to adhere to any restrictions imposed by the model. Given that EVM is usually the limiting factor in PA design under modulated conditions, the objectives, namely $P_{out,avg}$, $PAE_{avg}$, and the modulated average junction temperature ($T_{j,avg}$) were extracted at a fixed EVM of 8%. This EVM value represents the highest tolerable EVM for a 64-QAM modulation scheme. On the other hand, ACPR and Gain are used as constraints instead of objectives, where any design with an ACPR of more than $-30$ dBc (i.e., ACPR requirement for NR devices [31]) or less than 7 dB of Gain is deemed infeasible.

### B. LNA TRANSISTOR SIZING DESIGN VARIABLES AND OBJECTIVES

The operating parameter and design variables used for LNA transistor sizing are the same as those used for PA transistor sizing. However, the operating parameter range differs since the application targets low-noise amplification [6]. The number of designs has also significantly increased, totaling 2,352. This is a greater number because simulating small-signal objectives in the context of LNA design is generally faster. Table 3 outlines the operating parameter range (with linear spacing) corresponding to the $V_{GS}$ used in LNA transistor sizing, which translates to 90 − 375 mA/mm. Additionally, Table 4 summarizes the type, range, step, and number of points (with linear spacing) for the five design variables used for LNA transistor sizing.

**TABLE 3.** Operating parameter range for LNA optimization.

| Parameter | Range | Step | # Points |
|-----------|-------|------|----------|
| $V_{GS}$ (V) | $[-1.6, -1.0]$ | 0.1 | 7 |

Unlike in PA transistor sizing, self-heating is not included in the simulations, which significantly reduces computation time and increases the number of designs. A frequency of 30 GHz was selected once again, targeting 5G frequency bands. We primarily considered four objectives: $f_{max}$, Gain (at 30 GHz), and minimum noise figure (NF$_{min}$, at 30 GHz).

**TABLE 4.** Design variables for LNA optimization.

| Variable | Type | Range | Step | # Points |
|----------|------|-------|------|----------|
| $V_{DS}$ (V) | Float | [16, 28] | 2 | 7 |
| $N_f$ | Even Integer | [2, 8] | 2 | 4 |
| $W_f$ ($\mu$m) | Float | [25, 100] | 12.5 | 7 |
| GDG ($\mu$m) | Float | [22, 52] | 30 | 3 |
| GSG ($\mu$m) | Float | [33, 78] | 30 | 4 |

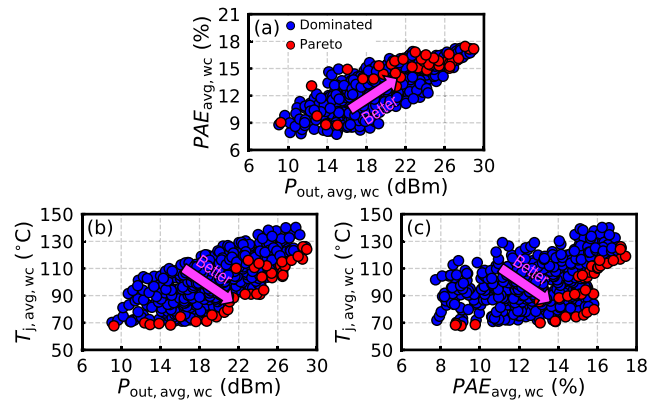

**FIGURE 2.** Two-dimensional cross-sections of the optimal trade-off surface showcasing the trade-off between (a) $PAE_{avg,wc}$ vs. $P_{out,avg,wc}$, (b) $T_{j,avg,wc}$ vs. $P_{out,avg,wc}$, and (c) $T_{j,avg,wc}$ vs. $PAE_{avg,wc}$. The Pareto optimal designs are highlighted in red, and the dominated designs are in blue.

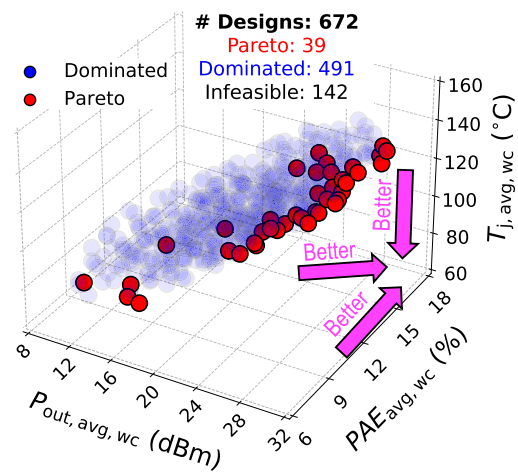

**FIGURE 3.** Optimal trade-off surface showcasing the trade-off between $T_{j,avg,wc}$ vs. $PAE_{avg,wc}$ vs. $P_{out,avg,wc}$. The Pareto optimal designs are highlighted in red, and the dominated designs are in blue.

### C. OPTIMAL TRANSISTOR SIZING FOR PAs

To find optimal device sizes in the context of PA applications, 4,704 SPICE simulations are done, accounting for 672 device designs and seven different bias conditions. All of the simulations in the present work were conducted using an Intel i9-9900 CPU (3.1 GHz). On average, each simulation takes approximately three minutes, resulting in approximately 9.8 days of simulation time to solely simulate these device designs under modulated conditions. After considering the worst-case performance for each objective, along with the imposed constraints (i.e., ACPR < −30 dBc and Gain > 7 dB) across the 7 bias conditions (to account for robustness), this number reduces to 530 robust designs, where 142 points are deemed infeasible (i.e., they do not meet the ACPR or Gain constraints). In this context, a design is considered robust if it accounts for the worst-case performance of all objectives within the given range of the operating parameter $V_{GS}$.

From these robust designs, the robust Pareto optimal designs are then found by maximizing $P_{out,avg,wc}$ and $PAE_{avg,wc}$ while minimizing $T_{j,avg,wc}$. This results in a total of 39 robust Pareto optimal designs. Fig. 2 showcases the 2D cross-sections of the optimal trade-off surface for various objectives, highlighting the Pareto optimal designs in red and the dominated designs in blue. In a similar manner, Fig. 3 shows the optimal trade-off surface, where the Pareto optimal designs are highlighted in red while the dominated designs are highlighted in blue. Here, each design corresponds to a unique transistor size, representing one specific combination of the five possible design variables. Based on these results, it can be seen that a variety of optimal designs with different trade-offs can be chosen. For example, if $P_{out,avg}$ is a top priority, then a robust Pareto optimal design can be selected with the desired $P_{out,avg}$ value at the expense of a higher $T_{j,avg}$, which is generally the case since a larger device is needed.

### D. OPTIMAL TRANSISTOR SIZING FOR LNAs

The optimal device sizes for LNAs are found using a procedure similar to that used for the sizing of PAs. However, 16,464 SPICE simulations were performed, accounting for 2,352 device designs and seven $V_{GS}$ bias conditions. These simulations were much faster, taking less than an hour to complete, given that these are small-signal simulations. When considering the worst-case performance for each objective, there are 2,352 robust designs. The robust Pareto optimal designs were then found by maximizing $f_{max,wc}$ and $Gain_{wc}$, and minimizing $NF_{min,wc}$, resulting in a total of 60 robust Pareto optimal designs. These results are summarized in the 2-D cross-sections of the optimal trade-off surface for the three objectives in Fig. 4, where the Pareto optimal and the dominated points are highlighted in red and blue, respectively. The optimal trade-off surface is also shown in Fig. 5 with the Pareto optimal designs highlighted in red and the dominated designs in blue.

## IV. DERIVATIVE-FREE OPTIMIZATION IN TRANSISTOR SIZING

In a variety of engineering optimization problems, the objective functions or constraints generally involve complex, time-intensive simulations, which is the case in the present work. In some instances, the objective function might not be differentiable or is too computationally expensive to differentiate. An elegant solution in such scenarios is to resort to derivative-free optimization, a black-box approach that does not require gradient information about the objective function [33], [34], [35], [36]. Derivative-free optimization particularly shines in the context of ML, more specifically, in hyperparameter tuning, where the objective is to find the set of hyperparameters that yields the best model performance [14], [37], [38], [39]. There are numerous open-source HPO frameworks that employ DFO methods, which include Autotune [37], HOLA [14], Hyperopt [38], and Optuna [39], to name a few. In this work, we have chosen to use HOLA. Although the selection of a specific HPO framework is not critical for our task, HOLA offers distinct advantages when addressing multiple objectives. It minimizes an overall scalar cost function by combining multiple objectives with a target-priority-limit scalarizer, streamlining the process of selecting a Pareto design.

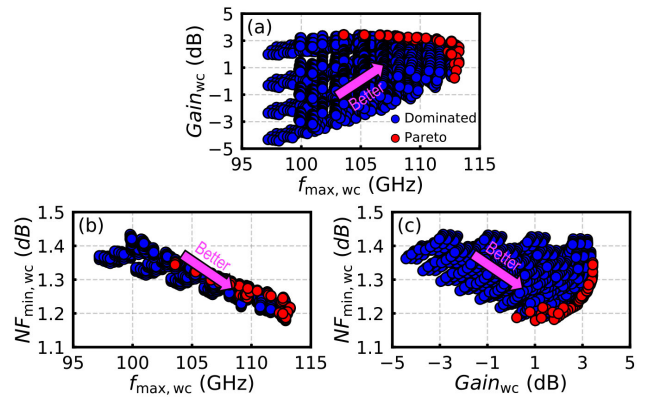

**FIGURE 4. Two-dimensional cross-sections of the optimal trade-off surfaceshowcasing the trade-off between (a) $Gain_{wc}$ vs. $f_{max,wc}$, (b) $NF_{min,wc}$ vs. $f_{max,wc}$, and (c) $NF_{min,wc}$ vs. $Gain_{wc}$. The Pareto optimal designs are highlightedin red, and the dominated designs are in blue.**

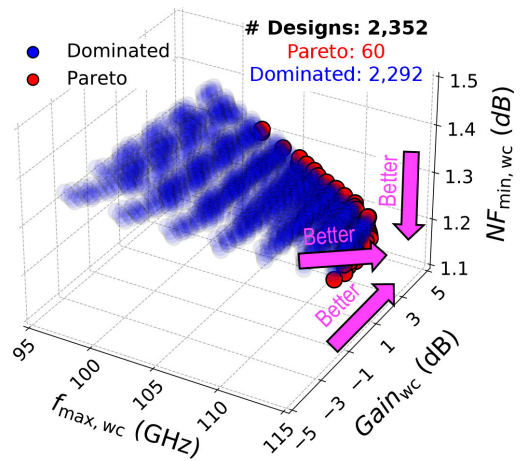

**FIGURE 5. Optimal trade-off surface showcasing the trade-off between $NF_{min,wc}$ vs. $Gain_{wc}$ vs. $f_{max,wc}$. The Pareto optimal designs are highlightedin red, and the dominated designs are in blue.**

With respect to transistor sizing optimization, we are presently dealing with five design variables, and our objectives for PA optimization are to maximize $P_{out,avg}$ and $PAE_{avg}$ while minimizing $T_j$. For LNAs, we are primarily interested in maximizing $f_{max}$ and Gain, and minimizing $NF_{min}$. Resource-intensive simulations are needed to compute these objectives, particularly for simulations requiring a modulated stimulus (e.g., a 64-QAM signal). If we were to consider every possible combination of design variables, the computation time would become prohibitively expensive and time-consuming, leading to the curse of dimensionality. By using a DFO method, particularly a hyperparameter optimizer like HOLA, the optimization process is steered toward regions in the design space where robust Pareto designs are likely to be found. It also drastically reduces the number of required simulations, reducing simulation time and computational power. Moreover, the optimization problem can be quickly adjusted if the design objectives or constraints change. However, one drawback of this method is that we do not know if a global optimal design was achieved. This is because DFO methods employ stochastic or heuristic approaches for search space sampling, which may not fully traverse the search space, potentially resulting in convergence to local minima.

**TABLE 5.** Summary of the mean scores* from various experiments using HOLA for optimal transistor sizing in PAs.

| Exp. | $P_{out,avg}$ (dBm) | | | $PAE_{avg}$ (%) | | | $T_{j,avg}$ (°C) | | | Mean Score* / % of Runs Within 5% of Optimal Score | | | | | Optimal Score |
|---|---|---|---|---|---|---|---|---|---|---|---|---|---|---|---|
| | Tgt. | Lim. | Pri. | Tgt. | Lim. | Pri. | Tgt. | Lim. | Pri. | $n = 15$ | $n = 30$ | $n = 45$ | $n = 60$ | $n = 75$ | |
| (a) | 28 | 8 | 1 | 17 | 5 | 1 | 125 | 150 | 1 | 0.182 / 27% | 0.100 / 42% | 0.068 / 55% | 0.050 / 70% | 0.033 / 81% | 0 |
| (b) | 23 | 12 | 1 | 15 | 5 | 1 | 95 | 145 | 1 | 0.186 / 21% | 0.092 / 50% | 0.065 / 62% | 0.038 / 77% | 0.023 / 86% | 0 |
| (c) | 12 | 5 | 1 | 9 | 5 | 1 | 70 | 125 | 1 | 0.109 / 25% | 0.060 / 49% | 0.040 / 70% | 0.027 / 84% | 0.022 / 92% | 0 |
| (d) | 25 | 10 | 1 | 15 | 5 | 1 | 105 | 160 | 1 | 0.139 / 17% | 0.073 / 39% | 0.051 / 60% | 0.035 / 78% | 0.022 / 87% | 0 |

Our proposed method is also broadly applicable to other engineering applications that require optimizing one or multiple objectives with various design variables. Some examples include compact model parameter extraction [40] or device TCAD simulations. We have demonstrated its efficacy in the context of transistor sizing, where carrying out exhaustive simulations ensures the certainty of the results. This approach allows us to confidently identify the global optimum in design. However, it is important to highlight that, in practical settings, conducting exhaustive simulations is often impractical. The primary advantage of our method is that it can lead to the identification of optimal or near-optimal device designs, achieving the set objectives with far fewer simulations than a full grid search would require.

### A. DERIVATIVE-FREE OPTIMIZATION USING HOLA

HOLA considers a vector $x$ within a subset $\mathcal{X}$ of $\mathbb{R}^n$ to represent the selected hyperparameters from a feasible set $\mathcal{X}$. The $i$-th element of vector $x$ is given by $x_i$. In the context of the present work, $x$ denotes a vector of the design variables of interest (e.g., $V_{DS}$, $W_f$, $N_f$, etc.). As previously defined, $f_{wc}$ represents the vector-valued objective under worst-case conditions, $f_{wc}(x) = (f_{1,wc}(x), \ldots, f_{k,wc}(x))$. Each objective has a desired direction of optimization, meaning some are to be maximized and others minimized.

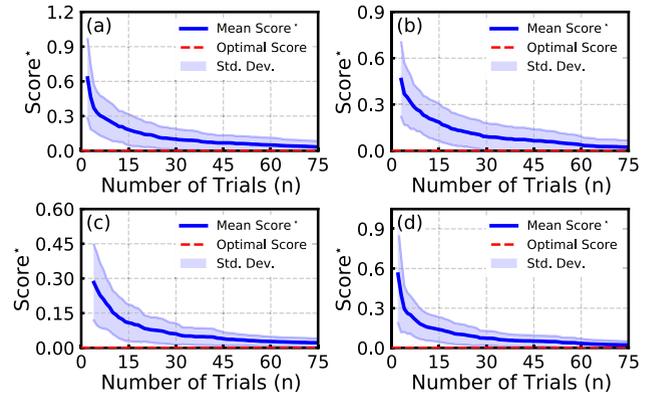

**FIGURE 6.** Plot of the Mean Score* as a function of the number of trials, showcasing four experiments with various targets, limits, and priorities for the objectives $P_{out,avg,wc}$, $PAE_{avg,wc}$, and $T_{j,avg,wc}$. Each experiment involves 100 independent simulation runs, with 75 trials in total per run. The plot illustrates the mean and standard deviation of Score* across runs, along with the optimal score for the corresponding experiment. The specific target (Tgt.), limit (Lim.), and priority (Pri.) values for each subfigure (a), (b), (c), and (d) are provided in Table 5.

The different choices of hyperparameters or design variables are compared by scalarizing the objectives into a singular cost function, $F_{wc}$, which we want to minimize. By convention, $F_{wc}(x) = +\infty$ designates an unacceptable hyperparameter vector. The transformation is realized through a target-limit-priority scalarization function $\phi$, which maps the vector of objectives to the real numbers. This can be expressed as $F_{wc}(x) = \phi(f_{wc}(x)) = \phi(f_{1,wc}(x), \ldots, f_{k,wc}(x))$. In essence, $F_{wc}(x)$ represents a traditional weighted sum of objectives.

The target-priority-limit scalarizer $\phi$ is separable and can be represented as:

$$\phi(u) = \sum_{i=1}^{k} \phi_i(u_i), \quad (2)$$

where each $\phi_i$ is characterized by three parameters: a target value $T_i$, a priority $P_i > 0$, and a limit $L_i$. For an objective function we want to minimize, the requirement is that $T_i \leq L_i$, and the function $\phi_i$ is defined as:

$$\phi_i(u_i) = \begin{cases} 0 & \text{if } u_i \leq T_i, \\ P_i \dfrac{u_i - T_i}{L_i - T_i} & \text{if } T_i \leq u_i \leq L_i, \\ +\infty & \text{if } u_i > L_i. \end{cases} \quad (3)$$

The scalarizer $\phi_i(u_i)$ is designed to be non-negative ($\phi_i(u_i) \geq 0$), achieving its minimum of zero when the objective is at or below the target value. When the objective value exceeds the target and stays below the limit, $\phi_i(u_i)$ will increase in a linear fashion, directly proportional to $P_i$. If $u_i$ exceeds the limit, it incurs an infinite penalty via the scalarizer function, i.e., $\phi_i(u_i) = \infty$. Essentially, the target represents the ideal or satisfactory performance level, the limit denotes the maximum allowable value, and the priority dictates the gradient between these bounds. In the case of an objective function we want to maximize, the inequality is inverted by setting $L_i \leq T_i$ and reversing the definition of $\phi_i(u_i)$.

**TABLE 6.** Summary of the mean scores* from various experiments using HOLA for optimal transistor sizing in LNAs.

| Exp. | $f_{max}$ (GHz) | | | Gain (dB) | | | $NF_{min}$ (dB) | | | Mean Score* / % of Runs Within 5% of Optimal Score | | | | | Optimal Score |
|---|---|---|---|---|---|---|---|---|---|---|---|---|---|---|---|
| | Tgt. | Lim. | Pri. | Tgt. | Lim. | Pri. | Tgt. | Lim. | Pri. | $n = 30$ | $n = 60$ | $n = 90$ | $n = 120$ | $n = 150$ | |
| (a) | 110 | 90 | 1 | 3 | 0.5 | 1 | 1.3 | 1.75 | 1 | 0.057 / 50% | 0.024 / 84% | 0.013 / 93% | 0.008 / 98% | 0.006 / 100% | 0 |
| (b) | 105 | 90 | 1 | 3.4 | 1 | 1 | 1.35 | 1.75 | 1 | 0.081 / 28% | 0.048 / 52% | 0.030 / 72% | 0.018 / 85% | 0.012 / 90% | 0 |
| (c) | 112 | 90 | 1 | 2.3 | 0.5 | 1 | 1.22 | 1.75 | 1 | 0.046 / 58% | 0.028 / 77% | 0.017 / 92% | 0.011 / 96% | 0.007 / 98% | 0 |
| (d) | 113 | 90 | 1 | 1.5 | 0.5 | 1 | 1.22 | 1.75 | 1 | 0.025 / 88% | 0.015 / 97% | 0.010 / 100% | 0.007 / 100% | 0.005 / 100% | 0 |

Having already established the scalarizer function $\phi$, the goal of HOLA is then to identify an optimal vector $x$ within $\mathcal{X}$ that minimizes $F(x)$ through simulations. We seek to solve the optimization problem:

$$\begin{aligned} \text{minimize} \quad & F_{wc}(x) \\ \text{subject to} \quad & x \in \mathcal{X}. \end{aligned} \quad (4)$$

In terms of sampling, a hybrid approach was employed, given the limitations of random sampling. The procedure is first started with random sampling, and subsequently, the sampling strategy is enhanced by incorporating Sobol sequences. This hybrid approach allows for a more systematic exploration of the hyperparameter space, which potentially improves the algorithm's performance. Once a baseline of hyperparameter values is established, a threshold is introduced. Beyond this threshold, HOLA employs a learned model designed to identify the most promising hyperparameter values with greater efficacy. A Gaussian Mixture Model (GMM) is then used to collect new hyperparameter values and concurrently present the findings [14].

At its core, HOLA begins by exploring a wide spectrum of promising hyperparameter values in order to identify regions in the hyperparameter space with the most potential. Subsequently, the GMM is tailored to align with the top 20–30% best-performing hyperparameter samples observed thus far. This allows the GMM to progressively detect and adjust to the distribution of best-performing hyperparameter values. As more samples are collected, the GMM's precision is enhanced, suggesting increasingly superior hyperparameter samples.

### B. HOLA SIMULATIONS FOR OPTIMAL TRANSISTOR SIZING OF PAs

We resort to HOLA primarily to recover the robust Pareto optimal designs. Doing so provides a better understanding of the trade-off space for more informed decision-making. Most importantly, HOLA allows us to find near-optimal or optimal designs in a relatively small number of trials as opposed to performing a more comprehensive search.

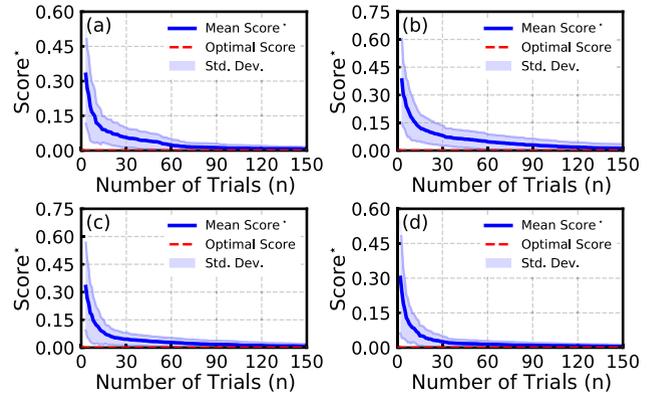

**FIGURE 7.** Plot of the Mean Score* as a function of the number of trials, showcasing four experiments with various targets, limits, and priorities for the objectives $f_{max,wc}$, Gain$_{wc}$, and NF$_{min,wc}$. Each experiment involves 100 independent simulation runs, with 150 trials in total per run. The plot illustrates the mean and standard deviation of Score* across runs, along with the optimal score for the corresponding experiment. The specific target (Tgt.), limit (Lim.), and priority (Pri.) values for each subfigure (a), (b), (c), and (d) are provided in Table 6.

This process begins by first setting the attributes of the design variables (or hyperparameters). We closely follow the description provided in Table 2 for the type, range, step, and number of points (given by a finite set of values) for optimal transistor sizing of PAs. Four different experiments are then considered, corresponding to various targets, priorities, and limits for the three objectives $P_{out,avg,wc}$, PAE$_{avg,wc}$, and $T_{j,avg,wc}$. The optimal score is then calculated for these four experiments based on the definition of $\phi$, which is given by (3) and represents the lowest possible score that can be obtained based on the given targets, priorities, and limits. For each of the four experiments, we perform 100 independent simulation runs using HOLA. Each run consists of 75 trials, during which we record the best score achieved to date, denoted as Score*. From the 100 independent runs, we calculate the mean and standard deviation of Score*. The results are summarized in Table 5 and Fig. 6, showing the Mean Score* for various number of trials ($n$) along with the corresponding targets (Tgt.), limits (Lim.), priorities (Pri.), and the optimal scores for the three objectives. In Table 5, a cell corresponding to a given experiment at trial number ($n$) is highlighted in green if more than 70% of the 100 independent runs achieve a Score* within 5% of the optimal score.

Based on these results, we found that after 60 trials, 70% of the independent runs are within 5% of the optimal score, which corresponds to searching less than 9% of the 672 robust designs. This outcome highlights the effectiveness of the derivative-free approach in identifying near-optimal or optimal designs with significantly fewer trials, as opposed to an exhaustive search among all 672 designs.

### C. HOLA SIMULATIONS FOR OPTIMAL TRANSISTOR SIZING OF LNAs

Similar to what was done for PAs, we resort once again to HOLA, but this time, a much larger number of designs is used (i.e., 2,352 robust designs as opposed to 672). We consider four distinct experiments with different targets, limits, and priorities for the three objectives, namely $f_{max,wc}$, $Gain_{wc}$, and $NF_{min,wc}$. For each of the four experiments, we conduct 100 independent simulation runs using HOLA. Each run consists of 150 trials, during which we record Score*. From these 100 independent runs, we compute the mean and standard deviation of Score*.

The results are summarized in Table 6 and Fig. 7, along with the corresponding targets (Tgt.), limits (Lim.), priorities (Pri.), and the optimal scores for the three objectives: $f_{max,wc}$, $Gain_{wc}$, and $NF_{min,wc}$. In addition, each cell representing a specific experiment at trial number $n$ is highlighted in green if more than 70% of the 100 independent runs achieve a Score* within 5% of the optimal score. These results showcase HOLA's effectiveness in finding near-optimal or optimal designs in less than 90 trials, corresponding to less than 4% of the total number of robust designs. We observe that as the number of designs increases, HOLA's capabilities become more useful in finding good designs in fewer trials with respect to the total number of designs.

In this particular example, HOLA evaluated less than 90 designs instead of going through all 2,352 designs. This approach is highly beneficial in problems with larger design spaces, such as in device-level TCAD simulations, which involve a greater number of design variables and lead to a vast number of design possibilities (in the order of millions). Despite the uncertainty of not knowing if the best possible (global) design has been achieved, there is a high likelihood that a near-optimal design has been identified with just a fraction of the simulations than what is required from a full-grid search.

## V. CONCLUSION

In this work, we have shown a robust Pareto design approach for sizing GaN HEMTs in the context of both PA and LNA settings. Five key design variables were addressed to target three different objectives. The device designs were assessed based on their worst-case performance over various values of $V_{GS}$. A derivative-free optimization method was then used to identify robust Pareto optimal designs to enhance our understanding of the trade-offs involved. While our DFO method does not guarantee a globally optimal design, it found Pareto optimal designs much faster than if we did a full-grid search. This approach provides a practical solution for sizing GaN HEMTs in 5G technology and is broadly applicable to various engineering design scenarios where multiple design variables and one or more objectives are at stake. Lastly, to support further research and implementation, we have made the code for this work available on our GitHub repository.[1]

---

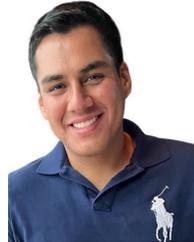

**RAFAEL PEREZ MARTINEZ** (Graduate Student Member, IEEE) received the B.S. degree (Hons.) in electrical engineering and the M.S. degree in electrical and computer engineering from Georgia Institute of Technology, Atlanta, GA, USA, in 2016 and 2019, respectively, and the M.S. and Ph.D. degrees in electrical engineering from Stanford University, Stanford, CA, USA, in 2023 and 2024, respectively.

He is currently a Research and Development Hardware Engineer with Broadcom Inc., San Jose, CA, USA. His research interests include semiconductor modeling, integrated circuit design, optimization, and machine learning.

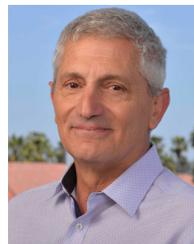

**STEPHEN BOYD** (Life Fellow, IEEE) received the A.B. degree in mathematics from Harvard University, Cambridge, MA, USA, in 1980, and the Ph.D. degree in electrical engineering and computer science from the University of California at Berkeley, CA, USA, in 1985.

He is currently the Samsung Professor in engineering and a Professor in electrical engineering with Stanford University, Stanford, CA, USA. His current research interests include convex optimization applications in control, signal processing, machine learning, and finance. He is a member of U.S. National Academy of Engineering (NAE), a Foreign Member of Chinese Academy of Engineering (CAE), and a Foreign Member of the National Academy of Engineering of Korea (NAEK).

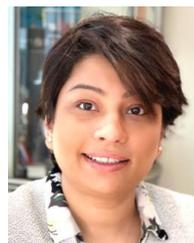

**SRABANTI CHOWDHURY** (Fellow, IEEE) received the M.S. and Ph.D. degrees in electrical and computer engineering from the University of California at Santa Barbara, Santa Barbara, CA, USA, in 2008 and 2010, respectively.

She is currently an Associate Professor in electrical engineering and a Senior Fellow of the Precourt Institute for Energy and Materials Science and Engineering (by courtesy), Stanford University. She demonstrated the first vertical power-switching transistor in GaN, known as CAVET. She specializes in wide bandgap and ultra-wide bandgap materials and device engineering, with a focus on energy-efficient system architecture and thermal management. She received the 2023 Technical Excellence Award from SRC for her work on diamond integration with GaN and SiC, the 2020 Alfred P. Sloan Fellowship in Physics, and the 2016 Young Scientist Award at the International Symposium of Compound Semiconductors (ISCS) for developing vertical GaN transistors. She actively serves on various IEEE committees, including EDS, VLSI, and IEDM.

● ● ●